# Evidence for mechanical softening-hardening dual anomaly in transition metals from shock compressed vanadium

Hao Wang[1, 2†], J. Li[1†], X. M. Zhou[1†], Y. Tan[1], L. Hao[1], Y. Y. Yu[1], C. D. Dai[1], K. Jin[1], Q. Wu[1], Q. M. Jing[1], X. R. Chen[2*], X. Z. Yan[3], Y. X. Wang[4], Hua Y. Geng[1, 5*]

[1] *National Key Laboratory of Shock Wave and Detonation Physics, Institute of Fluid Physics, CAEP, P.O. Box 919-102, Mianyang 621900, Sichuan, People's Republic of China*

[2] *College of Physics, Sichuan University, Chengdu 610065, People's Republic of China*

[3] *Jiangxi University of Science and Technology, Ganzhou 341000, Jiangxi, People's Republic of China*

[4] *College of Science, Xi'an University of Science and Technology, Xi'an 710054, People's Republic of China*

[5] *Center for Applied Physics and Technology, HEDPS, and college of Engineering, Peking University, Beijing 100871, People's Republic of China*

**Abstract:** Solid usually becomes harder and tougher under compression, and turns softer at elevated temperature. Recently, compression-induced softening and heating-induced hardening (CISHIH) dual anomaly was predicted in group VB elements such as vanadium. Here, the evidence for this counterintuitive phenomenon is reported. By using accurate high-temperature high-pressure sound velocities measured at Hugoniot states generated by shock-waves, together with first-principles calculations, we observe not only the prominent compression-induced sound velocity reduction, but also strong heating-induced sound velocity enhancement, in shocked vanadium. The former corresponds to the softening in shear modulus by compression, whereas the latter reflects the reverse hardening by heat. These experiments also unveil another anomaly in Young's modulus that wasn't reported before. Based on the experimental and theoretical data, we infer that vanadium might transition from BCC into two different rhombohedral (RH1 and RH2) phases at about 79GPa and 116GPa along the Hugoniot, respectively, which implies a dramatic difference in static and dynamic loading, as well as the significance of deviatoric stress and rate-relevant effects in high-pressure phase transition dynamics.

---

[†] *Contributed equally.*

[*] *Corresponding authors. E-mail: s102genghy@caep.cn, xrchen@scu.edu.cn*

# I. INTRODUCTION

Usually, compression increases not only the density, but also the mechanical modulus and sound velocity. The larger the mechanical modulus of a material becomes, the more it can resist external deformation[1, 2], a quality known as hardening. In contrast, increasing temperature often softens a material[3]. This phenomenon of compression-induced hardening and heating-induced softening (CIHHIS) is so general that it is considered as a *golden rule* in solid state physics, which had been well understood, and the variation in modulus can be modeled accurately.

Recently, this rule was predicted to fail in vanadium (V) and niobium (Nb). Both elements were predicted to exhibit compression-induced softening (CIS) in shear modulus $C_{44}$. Niobium manifests multiple softening at high pressures (HP), and a rhombohedral phase transition is induced in vanadium [4-7]. Another anomalous heating-induced hardening (HIH) in $C_{44}$ of both V and Nb was also predicted at high temperature (HT). Therefore, a CISHIH *dual anomaly* was predicted in V and Nb, both have a striking magnitude and unusual nature. Nevertheless, direct experimental evidence is still absent.

On the other hand, Suzuki *et al.*[6] and Landa *et al.*[7] once proposed the possibility of CIS on the phonon spectra of vanadium. This led to a hypothesis of concomitant structural transition driven by soft mode. The first-order nature of these transitions was elucidated by Wang *et al.*, who also found for the first time that $C_{44}$ of both V and Nb will increase at elevated temperature[4, 5]. This work unequivocally demonstrated that the softening-hardening anomaly in mechanical moduli is more fundamental than the accompanying structural transitions.

In experiment, X-ray diffraction (XRD) with diamond anvil cell (DAC) was used to detect the hypothesized BCC to rhombohedral (RH) transition in vanadium, in which the experimental transition pressure is still under debate[4, 8, 9]. The predicted reverse transition from RH back to BCC at HT was first predicted by Wang *et al.*, and was then confirmed by recent XRD and laser-heating DAC experiments[10, 11]. However, these static experiments detected only the RH1 phase[4, 8-12]. The RH2 is still *beyond the experimental scope*. The observed BCC-RH1-BCC transition is just the manifestation of the anomaly in electronic structure, rather than an evidence for the softening-hardening phenomenon itself. Direct evidence for the predicted dual mechanical anomaly[4, 5] is still required. It is crucial for understanding the properties of these metals at HP and HT.

In principle, the slope of acoustic phonon frequency at the long wave-length limit $\lim_{|k|\to 0} d\omega/dk$ can provide information about the elastic modulus and sound velocity. There have been several attempts to measure the phonon dispersion at HP for Ta, Nb, and V[13, 14], but the data quality is not accurate enough to derive useful elastic modulus or sound velocity. On the other hand, to measure the sound velocity directly by an ultrasonic method in the interested pressure range of 50~150GPa is very difficult.

Jing *et al.* tried to measure the lattice deformation magnitude under non-hydrostatic compression conditions using XRD and DAC, and the yield strength of Nb and Ta were then derived[15, 16]. Their data indeed revealed some anomalous behavior. However, since this method requires the elastic modulus as an input, the results cannot tell anything about the elastic modulus itself.

On the other hand, dynamic experiments with shock-waves and fast diagnostic techniques can be used to measure the longitudinal sound velocity ($C_l$). It had been employed to probe the solid-solid transitions and shock melting in many metals[17-23]. In vanadium, in addition to the shock Hugoniot[24-26], the jump in $C_l$ has been measured using transparent-window optical analyzer techniques to study the shock melting[27]. A similar method was also employed to detect the changes in yield strength[28]. Nevertheless, no studies have tried to probe the softening-hardening effect with dynamic experiments.

In this letter, a theoretical prediction of the CIS and HIH variation in the elastic modulus and sound velocity along the Hugoniot of vanadium is presented. It demonstrates how the dual anomaly manifests in dynamic shock conditions. Several already available shock-wave experiments were revisited to seek the possible signs of CISHIH, as well as their deficiency. Two new independent sets of shock-wave experiments were then carried out. They confirm the predicted CISHIH dual anomaly in both elastic modulus and sound velocity, and serve as the experiment support in vanadium.

## II. METHODS

### A. Elastic constants

The elastic constants were calculated using the energy-strain method[29, 30]. In general, the second elastic constants (SOEC) can be calculated based on the Euler strain ($\varepsilon$). In elastic theory, energy can be related to Euler strain through Taylor expansion in the form of strain tensor.

$$E = E_0 + V_0 \sum_i^6 \sigma_i \varepsilon_i + \frac{1}{2!} V_0 \sum_{i,j=1}^6 C_{ij} \varepsilon_i \varepsilon_j + \cdots \quad (1)$$

The coefficient of Taylor expansion is the elastic constant that needs to be solved.

$$C_{ij} = \frac{1}{V_0} \frac{\partial^2 E}{\partial \varepsilon_i \partial \varepsilon_j} \Big|_{\varepsilon=0} \quad (2)$$

The polycrystalline modulus are calculated from the single-crystal elastic constants by using the Voigt–Reuss–Hill (VRH) method[31, 32]:

$$9B_V = (C_{11} + C_{22} + C_{33}) + 2(C_{12} + C_{23} + C_{13}) \quad (3)$$

$$15G_V = (C_{11} + C_{22} + C_{33}) - (C_{12} + C_{23} + C_{13}) + 4(C_{44} + C_{55} + C_{66}) \quad (4)$$

$$B_R = \frac{1}{(S_{11} + S_{22} + S_{33}) + 2(S_{12} + S_{23} + S_{13})} \quad (5)$$

$$G_R = \frac{15}{4(S_{11} + S_{22} + S_{33}) - 4(S_{12} + S_{23} + S_{13}) + 3(S_{44} + S_{55} + S_{66})} \quad (6)$$

$$B_H = \frac{B_V + B_R}{2}, \quad G_H = \frac{G_V + G_R}{2} \quad (7)$$

$$E_{VRH} = \frac{9B_{VRH}G_{VRH}}{3B_{VRH} + G_{VRH}} \tag{8}$$

Where $S_{ij}=[C_{ij}]^{-1}$, called the compliance tensor. In which the quantity with subscript H (Hill) is what we reported in this paper. The symbol of $B$, $G$, and $E$ represent the bulk, shear, and Young's modulus, respectively.

In solid, since the compressional deformation and shear deformation of the material have non-vanishing stiffness, there are corresponding compression wave (longitudinal wave) and shear wave (transverse wave). Their velocity of propagation, namely, the sound velocity is evaluated according to their definition, which depends on the elastic properties of solid. For isotropic and homogeneous solid, the longitudinal and transverse sound velocities can be obtained as follows[33]:

$$C_l = \sqrt{\frac{B + \frac{4}{3}G}{\rho}}, \quad C_s = \sqrt{\frac{G}{\rho}} \tag{9}$$

Where $\rho$ is the density, and $B$ and $G$ represent the bulk and shear modulus of the solid, respectively. In a three-dimensional solid, the sound velocity of a body (bulk sound velocity) can be understood as an average of longitudinal sound velocity and shear sound velocity as:

$$C_b = \sqrt{C_l^2 - \frac{4}{3}C_s^2} \tag{10}$$

### B. Computational and experiment details

Theoretical calculation was carried out by using the VASP package based on density functional theory (DFT)[34, 35]. A plane-wave basis set was employed with a kinetic energy cutoff 900 eV. The electron-core interaction was described by a projector-augmented wave (PAW) pseudopotential[36]. The pseudopotential containing 13 valence electrons (including $3s^2$, $3p^6$, $3d^3$ and $4s^2$ states). Meanwhile, under high pressure, we discuss that the core-core overlap effect produced by this pseudopotential has no effect on the calculation results. The details can see

Supplemental Material (SM)[37] and also [38, 39]. The electronic exchange-correlation functional was set to the generalized gradient approximation (GGA) as parameterized by Perdew, Berke, and Ernzerhof (PBE)[40]. A Monkhorst-Pack (MP) grid with a size of 30×30×30 (24×24×24) was adopted for K-point sampling of the BCC (RH) structure, respectively. The self-consistent field (SCF) convergence tolerance was set as $10^{-8}$ eV per cell (0.001 eV/Å) for energy (force), respectively.

In order to disentangle the effect of compression and temperature, the elastic modulus and sound velocity of polycrystalline vanadium along both isotherm and Hugoniot paths were calculated. In the latter case the experimental Hugoniot data of Ref. [41] were used to set up the corresponding thermodynamic conditions. Thermal electronic contribution was fully taken into account via the electronic free energy functional of finite temperature DFT of Mermin[42]. Because the temperature range we are concerned with here is below the melting point, and previous studies have shown that in this temperature range the softening due to phonons (other than the corresponding thermal expansion contribution, which has been included in our work automatically) is much smaller than the electronic part[4, 43], we neglect this phonon contribution here to focus on the much more prominent electronic effects.

The shock-wave sound velocity was measured using a two-stage gas gun[44]. The poly-crystalline vanadium sample was purchased commercially with a purity of 99.9%. The sample thickness is 2.9 mm. For pressures below 117GPa, the direct-reverse impact method[44, 45] was employed. For higher pressures, a multiple-step method[46] with four steps was employed, with a step size of 0.5 mm. In all experiments, single crystal [100] LiF is used as the optical window for the Photonic Doppler Velocimetry (PDV) measurement. In this paper, we will focus mainly on the physical implications of these experimental data. The technical details, experimental sample information and data analysis of these experiments are presented in SM[37] and also [24, 47, 48].

## III. RESULTS AND DISCUSSION

Our calculations confirm the previously predicted softening-hardening effect in single-crystalline vanadium $C_{44}$[4]. The data at zero kelvin are in good agreement with other theoretical work[7, 12, 49], as shown in SM[37]. However, it is unclear whether this anomalous phenomenon still perceptible in polycrystalline samples or not. To examine this, the elastic modulus of polycrystalline vanadium at HP and HT along different isotherms were calculated using the VRH average of single-crystalline elastic constants.

The results are shown in Fig. 1(a), in which both the bulk modulus ($B$) and sound velocity ($C_b$) do not exhibit perceptible softening or hardening. However, a strong CISHIH is observed in the polycrystalline shear modulus ($G$) and Young's modulus ($E$) of vanadium [Figs. 1(b-c)]. The latter is in sharp contrast to the normally expected HIS behavior[50]. This softening-hardening effect modifies mainly the anisotropic components, but still presents in polycrystalline samples via the microscale anomalous single-crystal elastic response. The $B$ and $G$ reported by Rudd *et al.*[50] are in good agreement with our data. The small deviation in Fig. 1(c) at low pressure is due to the difference between the plane-wave PAW and FPLMTO method[7, 51], and the different convergence criteria that were employed.

The softening-hardening anomaly of $E$ in Fig. 1(b) has never been reported. We find it stems from the slight CIS in $C_{44}$ and $C_{11}$ of single-crystalline vanadium[4, 7, 52]. Its direct implication is that $C_l$ should also exhibit a strong softening-hardening effect. This insight paves the way to utilize shock-wave methods and polycrystalline samples to probe the evidence of CISHIH along the Hugoniot of vanadium. This kind of dynamic experiment is usually easier to carry out than static HP-HT experiments.

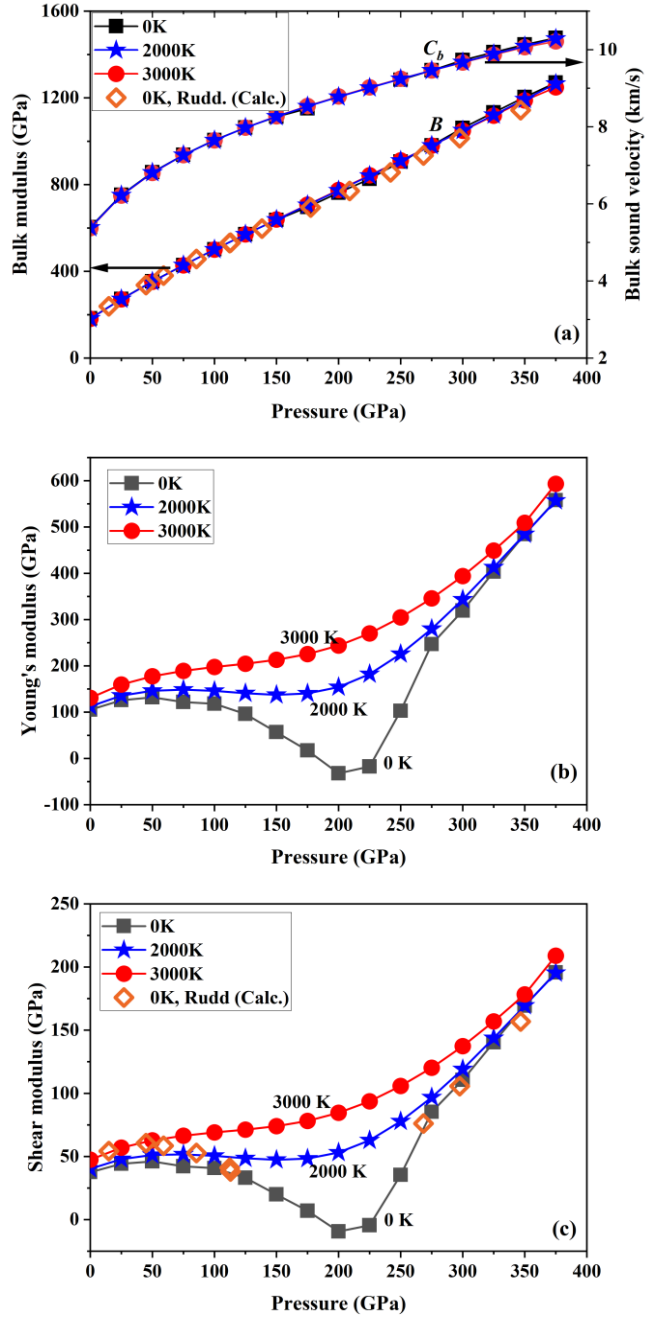

**Fig. 1.** (Color online) Calculated HP and HT elastic constants of polycrystalline BCC vanadium: (a) $B$ and $C_b$; (b) $E$; (c) $G$.

On the other hand, both theory and experiment suggest that vanadium transforms into RH phases and then back to BCC at HP[4-12]. This transition might alter the elastic constants and the corresponding sound velocity. Therefore, we also calculate the elastic modulus of polycrystalline RH phases, and compare them to BCC.

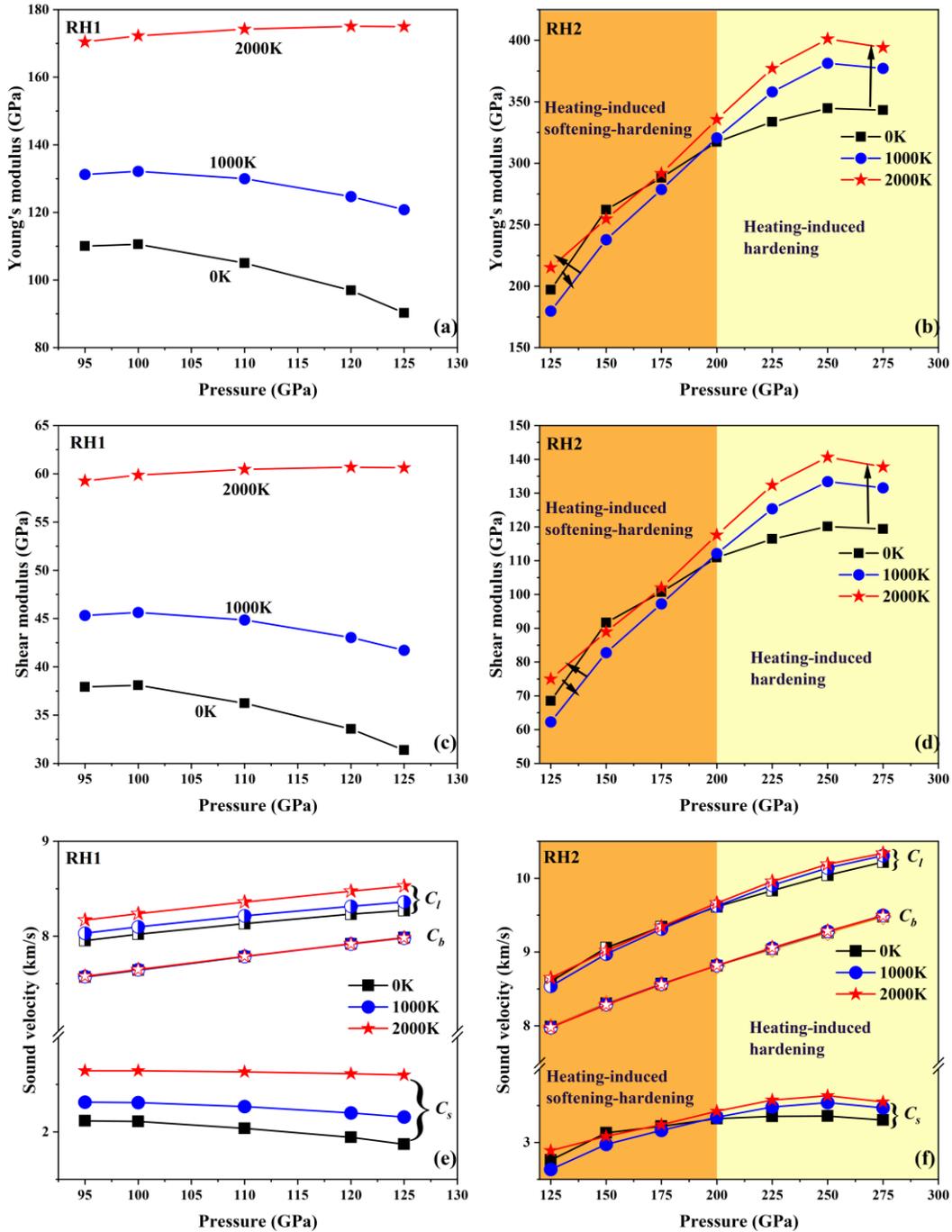

**Fig. 2.** (Color online) Calculated variation of polycrystalline elastic constants and sound velocities of RH1 and RH2 as a function of pressure at different temperatures, respectively. Notice the strong compression-induced softening and heating-induced hardening (CISHIH) in RH1 phase, and the HIS-HIH crossover (heating-induced softening to heating-induced hardening) in RH2 phase (as indicated by the arrows). In both phases, bulk sound velocity $C_b$ is almost temperature independent.

Figure 2 shows the polycrystalline elastic constants and sound velocity of RH1 and RH2 structure along isotherms, respectively. The calculation suggests that the bulk modulus of RH1 and RH2 are insensitive to temperature. It also should be noted that the bulk sound velocity of RH1 and RH2 are the same as that of BCC: all of them do not show any temperature dependence. The shear modulus in RH1 phase displays strong CIS and HIH (Fig. 2(c)). In terms of the sound velocity, the effect of HIH becomes more obvious (Fig. 2(e)). It should be noted that RH2 begins to show CIS at very high pressure (> 250 GPa), whereas the shear modulus still has HIH when pressure is greater than 200 GPa (Fig. 2(d)). On the other hand, if the pressure is less than 175 GPa, the shear modulus of RH2 softens first and then hardens up with the increased temperature (namely, heating-induced softening-hardening, as the arrows in Fig. 2(d) indicated). Of course, this situation mainly occurs when the temperature is low. As the temperature rises, RH2 still exhibits HIH. The Young's modulus (Fig. 2(b)) and longitudinal sound velocity (Fig. 2(f)) have similar behavior.

The predicted anomalous elastic modulus in BCC and RH phases implies that the softening-hardening effect might also be detectable in sound velocity. Our calculated sound velocity of $C_b$, $C_l$ and $C_s$ in polycrystalline vanadium along different loading paths are shown in Fig. 3 The main discoveries are: (1) $C_b$ increases monotonically with pressure, and is independent of temperature and structure; (2) $C_l$ and $C_s$ show CISHIH in both BCC and RH1 phase; (3) HIH also presents in the RH2 phase but is weaker; (4) BCC and RH phases have noticeable differences in their pressure dependence, which might be helpful for identifying the possible BCC→RH1→RH2 transitions[4, 12]; (5) there is a CIS anomaly in the $C_l$ of BCC at ~75GPa, a phenomenon that was not reported before.

The predicted softening up to ~100GPa in $C_s$ along the Hugoniot of BCC [Fig.3(b)] is striking. In addition, by comparing the isotherm with the Hugoniot line under the same pressure, both RH1 and RH2 have a larger $C_s$ if the temperature is higher. This makes their shock Hugoniot intersect with any given isotherms. That is, at a given pressure, the Hugoniot state always has a faster $C_s$ than the isotherm if the shock temperature is greater than the given isotherm, and vice versa, a distinctive signature of HIH anomaly. The same conclusion holds for $C_l$ in both BCC and RH phases.

Above analysis displays how the predicted softening-hardening effect will manifest in shock sound velocity, and provides a theoretical baseline to compare with experiment. Moreover, as shown in Fig. 3(b), the $C_s$ of both RH1 and RH2 along the Hugoniot depart from that of BCC. It therefore might be an indicator for the possible BCC-RH transition along shock Hugoniot.

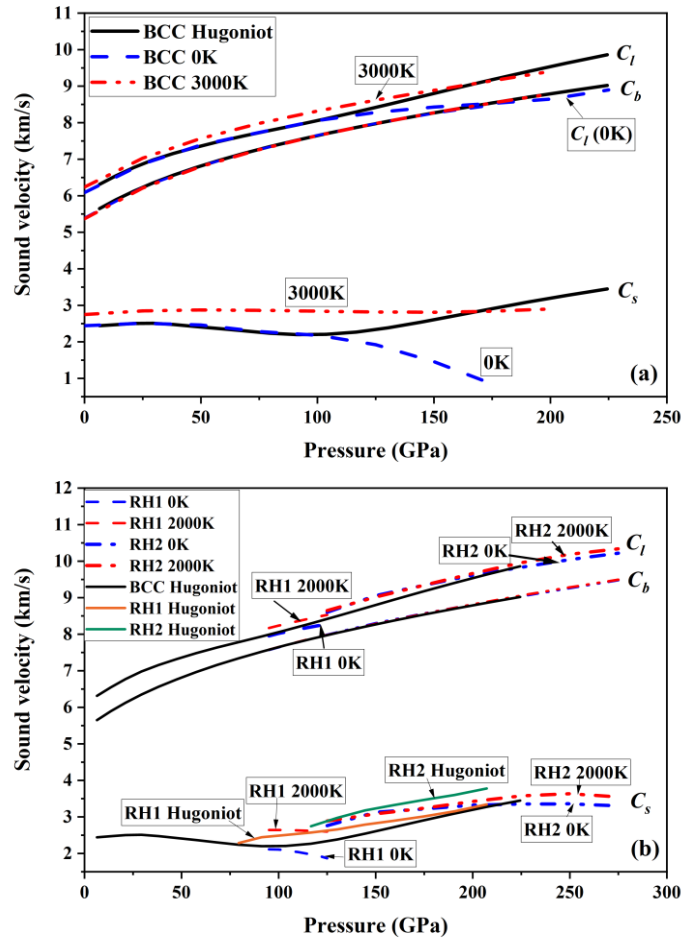

**Fig. 3.** (Color online) Calculated high-pressure high-temperature sound velocity of polycrystalline vanadium: (a) Comparison of Hugoniot with isotherms of 0K and 3000K in BCC phase; (b) Comparison of Hugoniot with isotherms of 0K and 2000K in RH phases, along with the Hugoniot of BCC phase.

With these advanced theoretical understandings, we revisit the published experimental data[27, 28]. Moreover, two new independent sets of experiments are also carried out, for the purpose to solidify the support of the CISHIH dual anomaly further. The experimental data are listed in Table I for the first data set that were carried out by X. M. Zhou *et al*. The second data set that were conducted independently by J. Li, Y. Tan *et al.* are listed in Table II. These two sets of experiments were performed with different equipment and diagnostic devices, for the purpose to account for the systematic uncertainty more appropriately, which was usually treated poorly in most experiments. The data of these four sets of shock experiments are shown in Fig. 4. It is evident that our theoretical data are in good agreement with the experimental data, especially for the BCC at low pressures.

**Table I**. Measured sound velocities of vanadium in DRI experiments. $\rho_0$ is the initial density of the flyer; $h_f$ and $W$ are the flyer thickness and velocity; $u_w$ is the particle velocity at sample/window interface; $t$ is the time duration of interface velocity plateau, $C_l$ is the Eulerian longitudinal sound velocity. The given uncertainties are the standard deviations.

| No. | $\rho_0$(g/cm$^3$) | $h_f$(mm) | $W$(km/s) | $t$ (ns) | $u_w$(km/s) | $P_s$(GPa) | $C_l$(km/s) |
|---|---|---|---|---|---|---|---|
| 1[a] | 6.084±0.023 | 2.920±0.002 | 0.94±0.01 | 905±2 | 0.631±0.006 | 9.99±0.20 | 6.56±0.13 |
| 2 | 6.084±0.023 | 2.971±0.002 | 2.89±0.02 | 827±2 | 1.886±0.010 | 38.40±0.80 | 7.06±0.14 |
| 3 | 6.067±0.023 | 2.946±0.002 | 4.02±0.02 | 742±2 | 2.588±0.014 | 59.07±0.12 | 7.54±0.15 |
| 4 | 6.104±0.023 | 2.950±0.002 | 4.71±0.02 | 708±6 | 3.012±0.017 | 73.40±0.15 | 7.68±0.39 |
| 5 | 6.071±0.023 | 2.962±0.002 | 5.10±0.03 | 682±2 | 3.251±0.018 | 81.89±0.16 | 8.02±0.16 |

[a] Elastic precursor wave correction was applied.

**Table II**. Measured sound velocities of vanadium in DRI and MSM impact experiments. $h_f$ and $W$ are the flyer thickness and velocity; $h_s$ is the sample thickness; $u_w$ is the particle velocity at sample/window interface; $t$ is the time duration of interface velocity plateau, $C_l$ is the Eulerian longitudinal sound velocity, $R$ is the catch-up ratio. The given uncertainties are the standard deviations.

| No.[a] | $h_f$(mm) | $W$(km/s) | $h_s$(mm) | $t$ (ns) | $R$ | $u_w$(km/s) | $P_s$(GPa) | $C_l$(km/s) |
|---|---|---|---|---|---|---|---|---|
| 1 | 0.919±0.004 | 6.319±0.032 | 1.018±0.004 | 134.4±6 | 4.365±0.336 | 3.996±0.040 | 173.0±1.9 | 9.261±0.364 |
|   |   |   | 1.517±0.004 | 112.4±6 |   |   |   |   |
|   |   |   | 2.027±0.004 | 89.0±6 |   |   |   |   |
|   |   |   | 2.515±0.004 | 67.5±6 |   |   |   |   |
| 2 | 0.926±0.004 | 5.924±0.030 | 1.016±0.004 | 138.9±6 | 4.445±0.356 | 3.762±0.038 | 157.7±1.7 | 9.106±0.362 |
|   |   |   | 1.514±0.004 | 109.6±6 |   |   |   |   |
|   |   |   | 2.026±0.004 | 94.1±6 |   |   |   |   |
|   |   |   | 2.518±0.004 | 71.6±6 |   |   |   |   |
| 3 | 2.929±0.004 | 6.498±0.032 | — | 610.7±6 | — | 4.123±0.041 | 116.2±2.1 | 8.481±0.355 |
| 4 | 0.770±0.004 | 6.398±0.032 | 1.211±0.004 | 183.2±6 | 6.446±0.474 | 4.311±0.04 | 193.1±1.4 | 9.744±0.501 |
|   |   |   | 1.708±0.004 | 155.7±6 |   |   |   |   |
|   |   |   | 2.224±0.004 | 131.1±6 |   |   |   |   |
|   |   |   | 2.708±0.004 | 111.2±6 |   |   |   |   |

[a] Shot Nos.1 and 2 were performed with vanadium flyer, and No.4 was performed with tantalum flyer to achieve a higher pressure. Shot No.3 was performed with direct-reverse impact set-up.

As mentioned above, $C_b$ is insensitive to temperature and structure. It therefore can be utilized as a standard reference to derive the "experimental" $C_s$ from the directly measured $C_l$ by using the relation $C_s^2 = 3(C_l^2 - C_b^2)/4$. The experimental $C_s$ (half-filled points in Fig. 4) match our DFT results very well, even though it was discovered that DFT systematically underestimates $G$ (as well as $C_s$) of vanadium at low pressures[52]. Some experimental data are slightly scattered, especially the point of Yu[28] at 31.8GPa. We re-scrutinized this datum, and found that it is flawed. The inappropriate elastic-wave correction undermines it slightly. We also note that the datum at ~150GPa of Dai departs significantly from all other points. Considering the obsolete technique employed by them, we re-check the validity of this point by carrying out a new experiment at almost the same pressure with a modern, advanced set-up and diagnostic methods. The new datum obtained at ~157GPa as shown in Fig. 4 is highly consistent with other data, but is much lower than Dai's original point. We therefore conclude that Dai's point at ~150GPa is invalid, and will exclude it in following analysis and discussion.

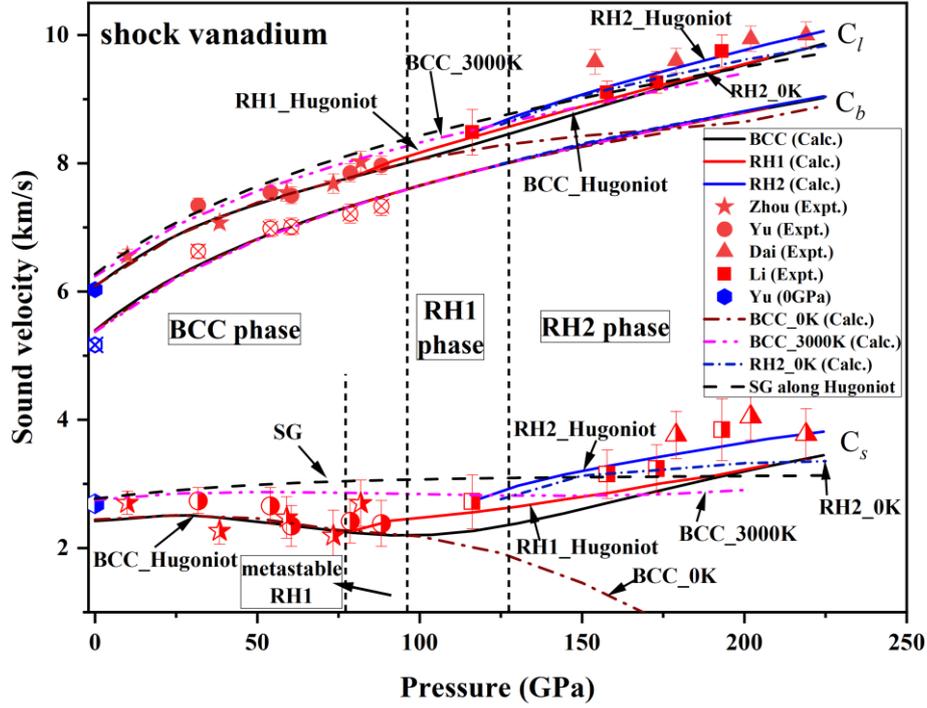

**Fig. 4.** (Color online) Calculated and experimental sound velocity of shocked polycrystalline vanadium along the Hugoniot. The shock experimental data of Dai[27], Yu[28] and our two new sets of experiments carried out by Zhou and Li are shown (solid points for $C_l$, half-filled points for $C_s$. The open crossed points are for $C_b$ as reported by Yu[28]), respectively. The isothermal curve of BCC at 0K and 3000K, and that of RH2 at 0K are also displayed for comparison. The Hugoniot sound velocity given by the Steinberg–Guinan (SG) model are also presented[50, 53]. The phase boundaries at 0K as estimated by Wang[4] are given as vertical dotted lines.

Clearly the experimental $C_s$ become smaller at higher shock pressures of up to ~100GPa. Within this pressure range, the shock temperature is low and its effect is insignificant, as shown in Fig. 4, which compares the shock Hugoniot with that of the BCC 0K isotherm. The observed softening in $C_s$ along the Hugoniot in Fig. 4 thus provides an evidence for CIS in vanadium. This softening is more obvious when comparing the high-pressure $C_s$ with that of zero pressure, after taking the uncertainty and scattering of all relevant experimental data into account. On the other hand, the $C_s$ of a normal metal usually becomes larger at HP, which is described well by the SG model[50, 53]. The applicability of the original SG model to vanadium is discussed (see SM[37] and also [54]). The CIS anomaly becomes more striking if compared to this normally expected HP behavior.

To illustrate this more clearly and intuitively, we plotted the deviation of the experimental and theoretical shear sound velocity with respect to the widely recognized SG model in Fig. 5(a). The shaded band in this figure contains all scattering and uncertainty of the experimental and theoretical data, and represents an overall variation trend of the shear sound velocity along the shocking path. It is obvious that when below a shock pressure of 100 GPa, not only the shear sound velocity at higher pressures becomes smaller than that of 0 GPa, but also its value continuously reduces with increasing pressure, and the deviation from the SG model at the same temperature becomes larger with the higher shock pressures. The largest softening takes place at ~75 GPa, with the shear sound velocity is reduced by more than 0.5 km/s.

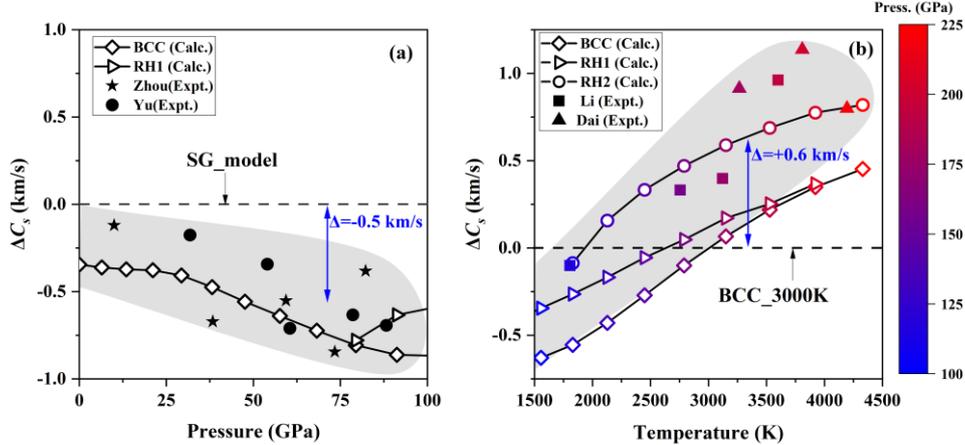

**Fig. 5.** (Color online) Comparison of the shock experimental and theoretical shear sound velocity at high-pressure and high-temperature with respect to those value of (a) the SG model with the same temperature along the shock path, and (b) along the 3000 K isotherm of BCC phase at the same pressure. The shaded bands contain the scattering and uncertainty of the experimental data, and represent an overall variation trend of the experimental data along the compression and heating path. The colormap in (b) indicates the shock pressure of each experimental or theoretical data.

When beyond 100GPa, the shock temperature effect becomes important, and HIH takes effect, as shown in Fig. 4 and Fig. 5(b). The calculated $C_s$ of shock Hugoniot of BCC departs from the 0K isotherm at ~100GPa, increasing to a higher value and intersecting the 3300K isotherm at ~180GPa. At this pressure, the experimental shear sound velocity is about 0.6 km/s faster than the 3000 K isotherm. The experimental $C_s$ indeed follows this trend, and continuously increases from ~75GPa to 200GPa. Both our newly obtained data and Dai's experiment show a distinguishable departure from the 0K isotherm of BCC phase. The magnitude of the departure continues to increase with pressure, and is larger than the experimental uncertainty and the scattering of the data. They provide an experimental support for an HIH anomaly in compressed vanadium, otherwise the experimental $C_s$ should be less than the 0K isotherm of BCC, which clearly is not the case. The same conclusion holds for $C_l$. Based on this observation, we infer that the $C_l$ and $C_s$ along an *isentropic* compression that is accessible using Z-machine or strong-laser technique, should lie between the Hugoniot and those of the 300K isotherm.

It should be pointed out that though the experimental $C_l$ and $C_s$ show an interesting HIH phenomenon, having a larger slope of $\frac{dC_l}{dP}$ and $\frac{dC_s}{dP}$ along Hugoniot than any isotherms when beyond 75GPa, as shown in Fig. 4, the experimental data in fact do not match the BCC phase. This is in a sharp contrast to that for 0~75GPa (in that range the experimental $C_l$ and $C_s$ are in good agreement with the BCC). It is known that in vanadium a structural transition to RH takes place when compressed at room temperature. For comparison, the sound velocity of both RH1 and RH2 along the Hugoniot are calculated and displayed in Fig. 4. It is evident that the $C_s$ and $C_l$ of RH1 are in better agreement with the experimental data in the pressure range of 79-116GPa, and the $C_s$ and $C_l$ of RH2 are in better agreement with the experimental data in the pressure range of 175-220GPa.

We noticed that around 116 GPa, our experimental values are consistent with both RH1 and RH2 phases, and the theoretical calculation values of RH1 and RH2 have an intersection. Therefore, we infer that the transition pressure of RH1-RH2 of vanadium may be 116GPa under shock loading. We also admit that more precise experimental measurements are needed to support our results within 116-175GPa. We thus hypothesize that there might have a BCC→RH1→RH2 transition in shocked vanadium. The estimated transition pressures ($P_t$) are 79GPa for BCC→RH1 and 116GPa for RH1→RH2.

The $P_t$ of 79GPa for BCC→RH1 is close to Ding's static DAC experimental result of 69GPa[8]. However, a $P_t$ as low as 30GPa was also reported[9], reflecting the challenge to address this problem experimentally. The discrepancies in the DAC experiments were largely attributed to non-hydrostatic conditions[9]. So far, equilibrium phase diagram of compressed vanadium was calculated only for hydrostatic conditions [4], and there is no attempt to take deviatoric stress into account. In this sense, the equilibrium and hydrostatic DFT results might underestimate the stability of RH if deviatoric stress is involved.

Dynamic shock compression always contains non-equilibrium and rate-dependent kinetic effects, which have not been included in any theoretical phase diagram. From the experimental data, we find that shocked vanadium has a sound velocity closer to RH phases rather than BCC when above 79GPa. The occurrence of these HP phases, which are slight distortions of the BCC structure, could be due to the dynamic, non-equilibrium and non-hydrostatic nature of planar shock-waves. However, it should be pointed out that as shown in Fig. 4, Zhou's data of $C_s$ are scattered slightly around the BCC Hugoniot and have large error bars when relative to the difference between the BCC and RH1 Hugoniot; Li's data cannot unequivocally distinguish between BCC, RH1, and RH2 at 115GPa, or between RH1 and RH2 at 165GPa. More accurate sound velocity data are definitely required to pin down the exact $P_t$, even though our current experiments are compatible with the hypothetical shock-driven BCC-RH phase transitions.

## IV. CONCLUSION

In summary, we predicted that the CISHIH dual anomaly in single-crystalline vanadium also presents in shocked polycrystalline vanadium. We also predicted an anomaly in both $E$ and $C_l$ that was not reported before. This theoretical insight paves the way to probing the softening-hardening anomaly in shock-wave experiments directly. We then carried out two new sets of shock-wave sound velocity measurements. These new data, together with previously reported experiments, provide an evidence for a CIS anomaly up to ~75GPa. The experiments also show significant HIH in shocked vanadium from 75-220GPa, where both $C_l$ and $C_s$ along the Hugoniot increase much faster than any isotherms. These observations further improve the understanding of the CISHIH double anomaly predicted in vanadium. [4-7, 12].

In addition, from the experimental and theoretical data, we also *infer* that shocked vanadium might experience a series of structural transitions: the first one at 79GPa for BCC→RH1, and the second one at 116GPa for RH1→RH2. These hypothesized transitions make vanadium harden further, and are fully compatible with all available experimental data. This conclusion highlights the elusive and highly sensitive HP-HT behavior of the group VB metals under dynamic loading. Our findings could stimulate further theoretical and experimental research on this prominent CISHIH dual anomaly for practical applications. It also calls for more accurate experiments to prove or disprove the hypothesized BCC-RH transitions in shocked vanadium.

**Acknowledgments**

This work was supported by the NSAF under Grant Nos. U1730248 and U1830101, the National Natural Science Foundation of China under Grant Nos. 11672274, 11872056, 11904282, 11704163 and 11804131, the CAEP Research Project under Grant No. CX2019002, the Science Challenge Project TZ2016001, the China Postdoctoral Science Foundation under Grant No. 2017M623064, the Natural Science Foundation of Jiangxi Province of China under Grant No. 2018BAB211007. The simulation was performed on resources provided by the Center for Comput. Mater. Sci. (CCMS) at Tohoku University, Japan.

## Author Contributions

H.Y.G. conceived and designed the project. H.W. performed the calculation. X.M.Z., J.L., Y.T., and L.H. performed the experiments. All authors analyzed the data. H.Y.G. and H.W. wrote the draft. All authors contributed to revise the manuscript.

# Supplementary Material

## Evidence for mechanical softening-hardening dual anomaly in transition metals from shock compressed vanadium

Hao Wang[1, 2†], J. Li[1†], X. M. Zhou[1†], Y. Tan[1], L. Hao[1], Y. Y. Yu[1], C. D. Dai[1], K. Jin[1], Q. Wu[1], Q. M. Jing[1], X. R. Chen[2*], X. Z. Yan[3], Y. X. Wang[4], Hua Y. Geng[2, 5*]

[1] *National Key Laboratory of Shock Wave and Detonation Physics, Institute of Fluid Physics, CAEP, P.O. Box 919-102, Mianyang 621900, Sichuan, People's Republic of China*

[2] *College of Physics, Sichuan University, Chengdu 610065, People's Republic of China*

[3] *Jiangxi University of Science and Technology, Ganzhou 341000, Jiangxi, People's Republic of China*

[4] *College of Science, Xi'an University of Science and Technology, Xi'an 710054, People's Republic of China*

[5] *Center for Applied Physics and Technology, HEDPS, and college of Engineering, Peking University, Beijing 100871, People's Republic of China*

## 1. Experimental details

The sound velocity (SV) at shocked states was measured using a 28 mm bore two-stage gas gun. The poly-crystalline vanadium sample was purchased commercially with a purity of 99.9%. For pressure below 117 GPa, the direct-reverse impact (DRI) method was employed. The DRI experimental configuration is illustrated in Fig. S1. The flyer, namely vanadium, impacted [100]LiF window directly. The impact velocity and particle velocity at the flyer/window interface were recorded with the Photonic Doppler Velocimetry (PDV). As shown in Fig. S1, the impact generates shock waves propagate into the flyer and the window, respectively. When reaching the rear surface of the flyer,

---

†*Contributed equally.*

*Corresponding authors. E-mail: s102genghy@caep.cn, xrchen@scu.edu.cn*

the shock wave reflects as a rarefaction wave, which unloads the shocked vanadium. The particle velocity $u_s$, shock velocity $D_s$, and Eulerian longitudinal sound velocity $C_l$ in the sample (*i.e.* the flyer) can be expressed as:

$$u_s = W - u_w \qquad (1.1)$$

$$D_s = \frac{\rho_{0w}(C_{0w} + \lambda_w u_w) u_w}{\rho_{0s}(W - u_w)} \qquad (1.2)$$

$$C_l = \frac{h_s}{\Delta t - h_s / D_s} \frac{\rho_{0s}}{\rho_s} \qquad (1.3)$$

where $W$ is the impact velocity, $\rho_{0s}$ and $\rho_s$ are the initial density and shock compressed density of sample, $\rho_{0w}$ is the initial density of window, $u_w$ is the measured particle velocity at the sample/window interface, $C_{0w}$ and $\lambda_w$ are the Hugoniot parameters of [100]LiF window, $h_s$ is the thickness of sample, $\Delta t = t_2 - t_1$. Here $t_1$ and $t_2$ are nearly equal to the impact time and the arrival time of the rarefaction wave at the sample/window interface, respectively.

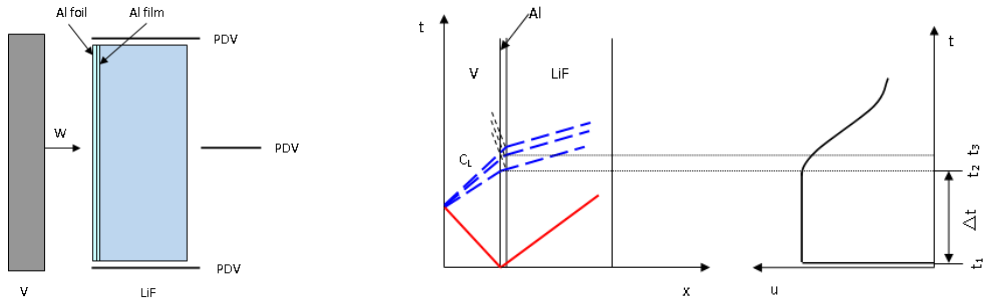

**Fig. S1** (Color online) Schematic for direct-reverse impact experimental set-up (left) and the corresponding wave interaction (right). Details of this experimental method also can be found in Ref. [1].

For higher pressures, the multiple-step method (MSM) with four steps was employed, with a step size of ~0.5 mm and 10 mm in diameter. The MSM experimental configuration based on rarefaction wave overtaking method is illustrated in Fig. S2. The details of rarefaction wave overtaking method for sound velocity measurement were described elsewhere [see, for example, Ref.[2]]. Four vanadium samples of different thickness (1–3 mm) formed a stepped wedge, and each sample was backed with an [100]LiF window. Tantalum or vanadium was used as the flyer at different shock pressure. The arrival time of rarefaction wave at the respective sample/window interfaces was determined by measuring the interfacial particle velocity profiles using PDV. According to the wave interactions as shown in Fig. S2, the Eulerian longitudinal sound velocity ($C_l$) in the sample can be expressed as:

$$\frac{1}{C_l}\frac{\rho_{0s}}{\rho_s} = \frac{1}{D_s} - \frac{1}{R}\left(\frac{1}{D_f} + \frac{1}{C_f}\frac{\rho_{0f}}{\rho_f}\right) \quad (1.4)$$

where $\rho_0$ and $\rho$ are the initial and compressed density, $D$ is the shock velocity, with the subscript $f$ and $s$ denote flyer and sample, respectively. $R$ is the catch-up ratio defined as $d_{max}/h_f$, here $d_{max}$ is the extrapolated thickness of sample of which the leading overtaking rarefaction wave from the rear side of the impactor exactly catch up the forward-going shock wave at the sample/window interface, and $h_f$ is the thickness of the flyer, respectively.

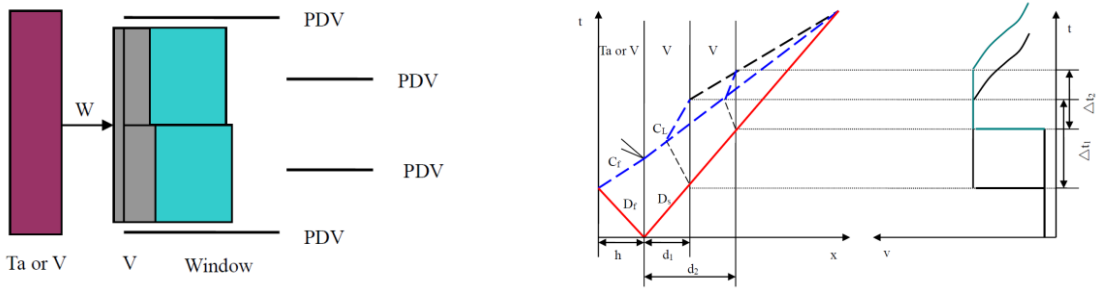

**Fig. S2** (Color online) Schematic for multiple-step impact experimental set-up (left) and the respective wave interactions and interfacial particle velocity profiles (right).

Typical recorded interfacial particle velocity profiles are shown in Fig. S3, from which the high-pressure sound velocity was derived. The experimental data are listed in Table I of main text for the first data set that were carried out by X. M. Zhou *et al*. The second data set that were conducted independently by J. Li, Y. Tan *et al.* are listed in Table II of main text. These two sets of experiments were performed with different equipment and diagnostic devices, for the purpose to account for the systematic uncertainty more appropriately, which was usually treated poorly in most experiments.

For the second set of experiments, a total of five DRI experiments were performed, with one at a single-stage gas gun and other four at a two-stage gas gun. Impact speed is measured by a magneto-coil system with an uncertainty of 0.5%. The Hugoniot states (shock wave speed, particle velocity, shock density and pressure, *etc.*) were determined for each experiment by the impedance-match method with measured impact speed and the known Hugoniot data of the vanadium sample and window materials. Uncertainty in the recorded release time is about 2 ns for most experiments. The fourth datum entry has an exception of 6 ns, due to the noises and unsteady quality presented in the wave profiles, which is the main source of uncertainty in sound velocity determination. For the lowest shock pressure (the first datum entry in Table 1), elastoplastic two-wave propagations occur and the reflection of the elastic precursor at the rear-surface of the sample (flyer) is taken into account in the calculation of the longitudinal sound velocity at the Hugoniot state.

The first set of experiments include both DRI and MSM methods. The samples were carefully selected and all have an initial density of $\rho_{0s} = 6.104$ g/cm$^3$. The adopted [100]LiF window for both sets of experiments has an initial density of $\rho_{0w} = 2.638$ g/cm$^3$, whose Hugoniot relation is $D_w$=5.148 km/s + 1.353$u_w$ [3]. The parameters for Ta flyer are[4, 5]: $\rho_0$=16.65 g/cm$^3$, and $D$=3.329 km/s + 1.307$u_w$.

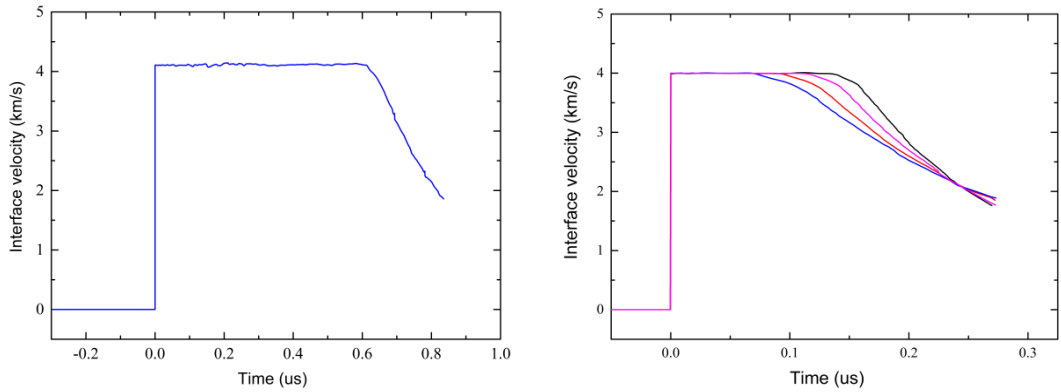

**Fig. S3** (Color online) Typical measured particle velocity profiles of vanadium in DRI (left) and MSM (right) impact experiments with a two-stage light gas gun and PDV at high pressure.

## 2. The structure of RH1 and RH2

According to existing research, both RH1 and RH2 belong to the trigonal crystal system. The unit cell structures of the two are shown in the figure. For the distinction between RH1 and RH2, the main reason is the relative size of the α angle to 109.47° (rhombohedral axes). When α<109.47°, the rhombohedral structure is called RH2; when α>109.47°, the rhombohedral structure is called RH1.

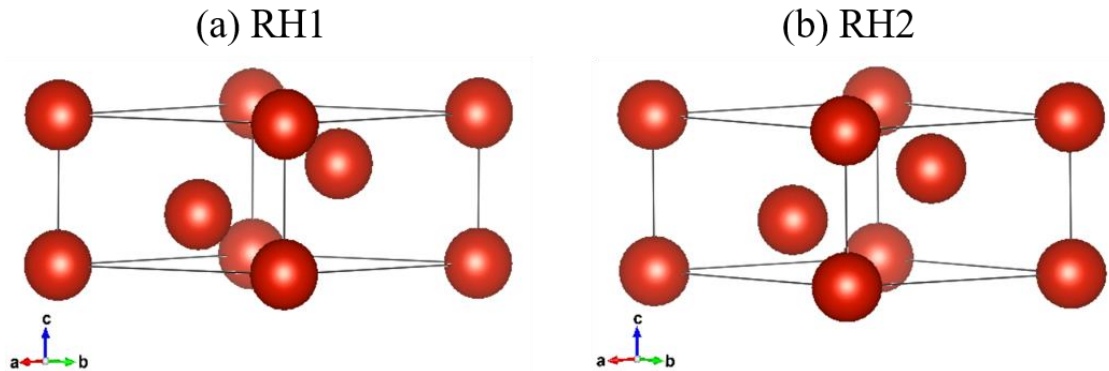

**Fig.** S4 (Color online) The structure of RH1 and RH2.

## 3. Elasticity along isotherms

The mechanical stability criterion formula[6] are as follows:

Cubic:

$$C_{11} > 0; \quad C_{44} > 0; \quad C_{11} > |C_{12}|; \quad (C_{11} + 2C_{12}) > 0 \tag{2.1}$$

Rhombohedral:

$$C_{11} > |C_{12}|; \quad C_{44} > 0$$
$$C_{13}^2 < \frac{1}{2}C_{33}(C_{11} + C_{12}) \tag{2.2}$$
$$C_{14}^2 < \frac{1}{2}C_{44}(C_{11} - C_{12}) = C_{44}C_{66}$$

With these criteria, we found that the stability region of BCC phase is in good agreement with others' work[7]; and RH1 (RH2) is stable in a pressure region of 75–125GPa (125–275 GPa), respectively.

In Fig. S5, the elastic constants of single-crystalline vanadium calculated at different pressure along several isotherms are given. It can be seen that the softening effect in $C_{44}$ under high pressure (HP) is weakened with increasing temperature, as shown in Fig. S5(a). In addition, the variation of $C_{44}$ with pressure as calculated by us is in good agreement with other theoretical results (Fig. S5(b))[7-10]. Figure S4(c) shows that elastic modulus (EM) $C_{11}$ softens at ~225 GPa (as marked by the arrow), which leads to $C'=(C_{11}-C_{12})/2$ also softening at the same pressure (Fig. S5(d)). From the anomalous variation in $C_{44}$ and $C'$, it can be deduced that polycrystalline vanadium should have a large compression-induced softening (CIS) when below 200 GPa and heating-induced hardening (HIH) when below 350 GPa in shear modulus.

Figures S6 and S7 display the change in elastic constants of vanadium in RH1 and RH2 structure, respectively. These two structures might be metastable in the given temperature-pressure range [14].

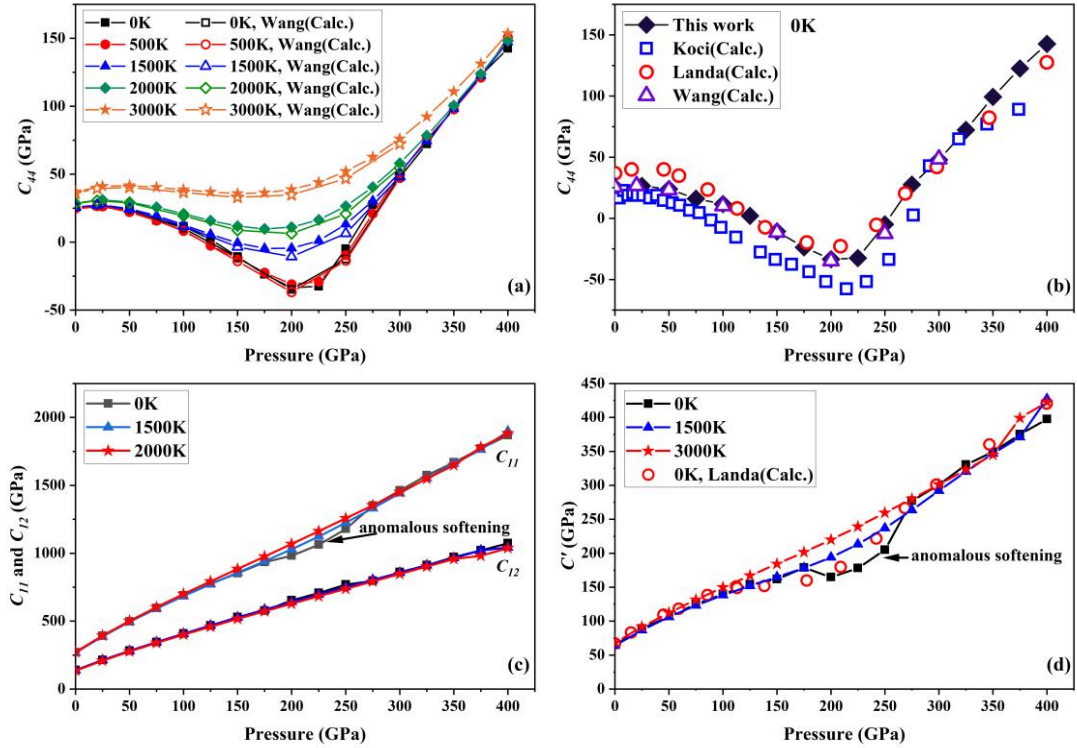

**Fig. S5** (Color online) Calculated variation of single-crystalline elastic constants of BCC structure of vanadium as a function of hydrostatic pressure at different temperature (filled symbols). The theoretical results of other work (open symbols) [7-10] are also plotted for comparison.

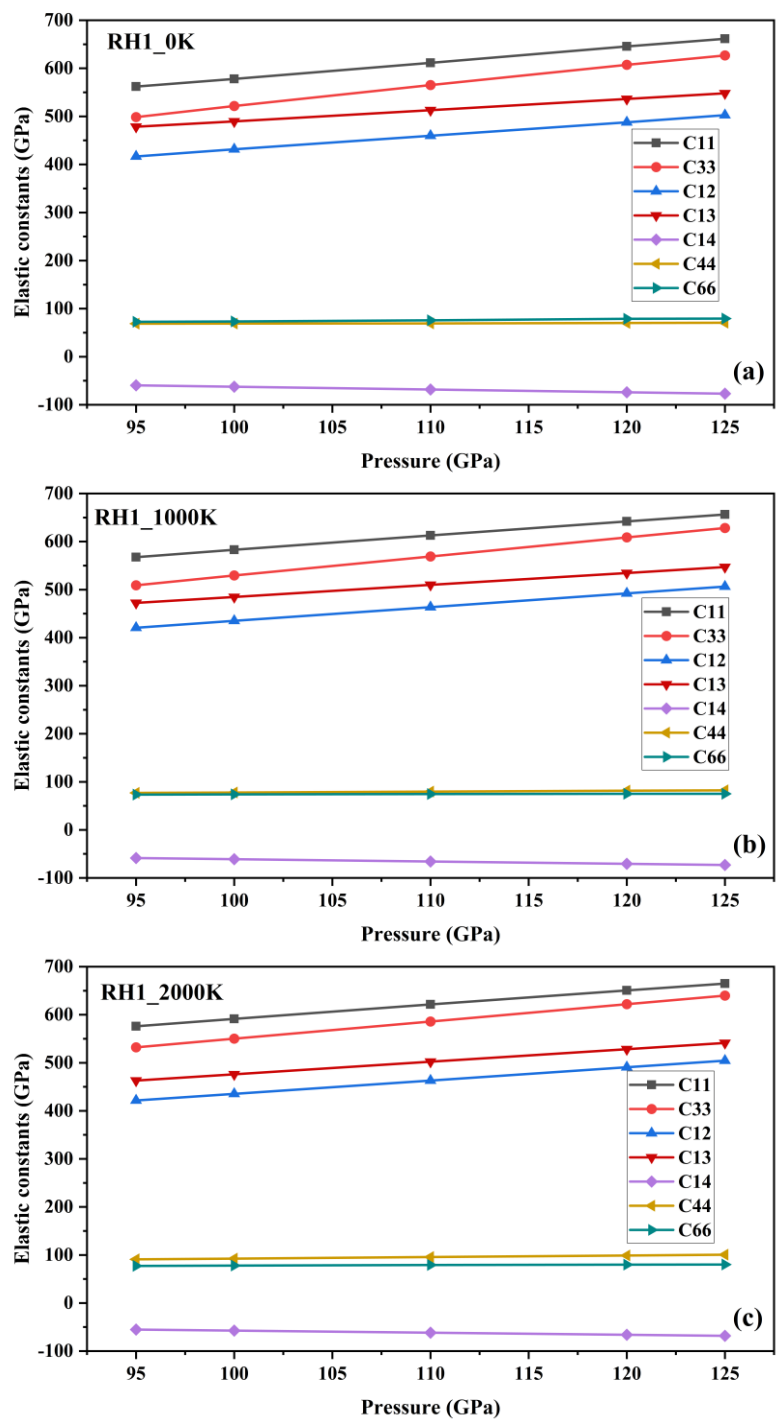

**Fig. S6** (Color online) Calculated variation of single-crystalline elastic constants of RH1 structure of vanadium as a function of pressure at different temperatures.

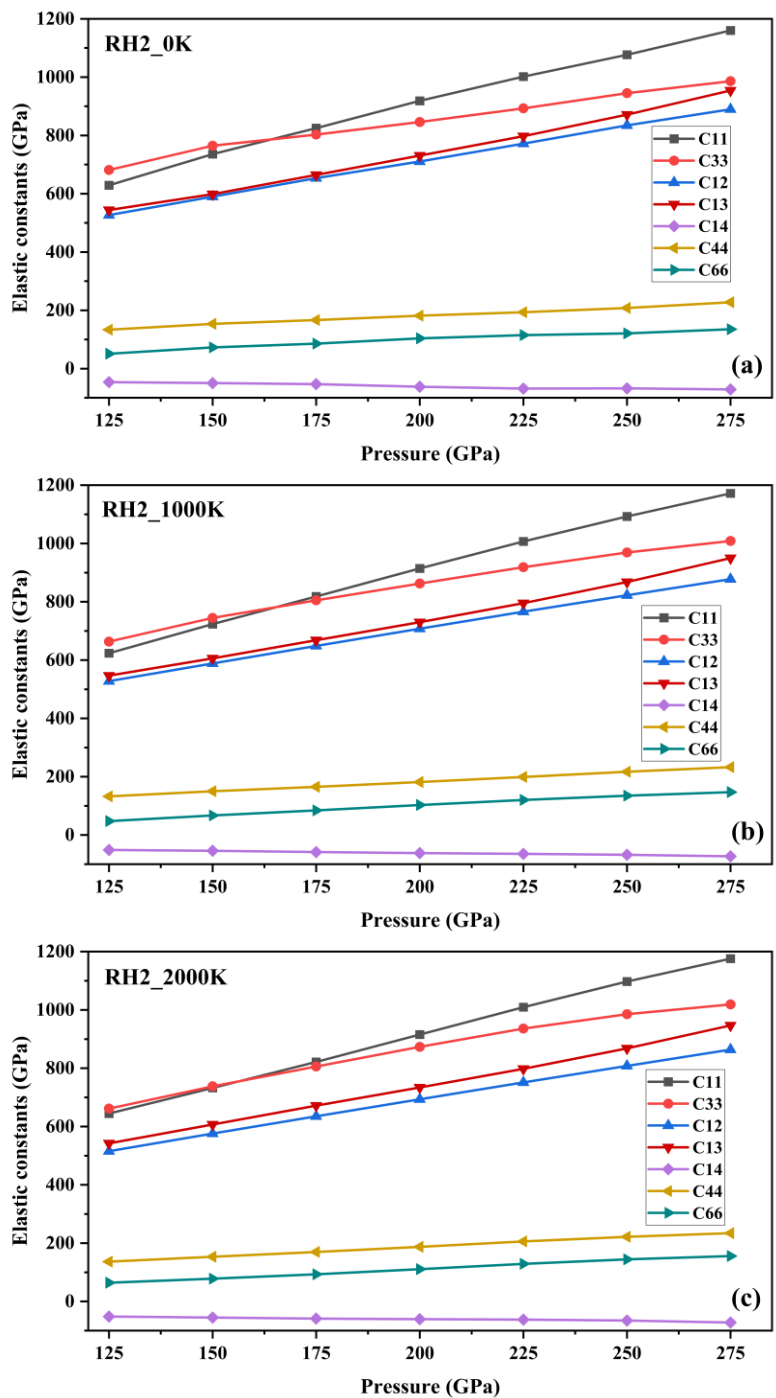

**Fig. S7** (Color online) Calculated variation of single-crystalline elastic constants of RH2 structure of vanadium as a function of pressure at different temperatures.

**Table S1 The experimental Hugoniot data for vanadium[11]**

| Vanadium | |
|---|---|
| Pressure (GPa) | Temperature (K) |
| 0 | 300 |
| 6.44 | 316 |
| 13.48 | 337 |
| 21.11 | 372 |
| 29.34 | 426 |
| 38.16 | 503 |
| 47.59 | 606 |
| 57.61 | 736 |
| 68.22 | 897 |
| 79.44 | 1089 |
| 91.25 | 1309 |
| 103.66 | 1557 |
| 116.66 | 1830 |
| 130.26 | 2128 |
| 144.46 | 2449 |
| 159.25 | 2790 |
| 174.64 | 3150 |
| 190.63 | 3528 |
| 207.22 | 3921 |
| 224.4 | 4329 |

## 4. Elasticity along shock Hugoniot

The variation of elastic moduli of vanadium in the structures of BCC, RH1 and RH2 along the adiabatic shock Hugoniot are calculated and given in Fig. S8. In these calculations, the experimental Hugoniot data ($P$, $T$) are taken from Ref. [11], which are listed in Table S1 for reference. As shown in Fig. S8, with increase of the shock pressure, the elastic constants $C_{11}$, $C_{12}$ and $C'$ of the BCC structure increase almost linearly, suggesting that BCC phase has a strong capability to resist vertical deformations. However, the elastic constant $C_{44}$ softens within a pressure range up to 110 GPa, which indicates that vanadium is weak in resistance to shear distortion, and the structure is prone to transversal deformation. This could be beneficial for the possible non-equilibrium phase transformation from BCC into RH phases, especially when in a deviatoric stress environment. However, this pressure range of softening along shock path is significantly narrower than that along the isotherms (e.g., at 0 K). The difference is mainly due to the influence of the simultaneous increasement of both temperature and pressure along the Hugoniot, in which the heating-induced hardening effect increases $C_{44}$ greatly. All single-crystalline elastic constants of RH1 and RH2 structures increase monotonously with the impact pressure, and the softening-hardening behavior (as shown in Fig. S7) cannot be directly observed in shocked RH phases.

Figure S9 displays the calculated shear modulus and Young's modulus of polycrystalline vanadium in BCC, RH1 and RH2 structures along the shock Hugoniot. Figure S9(a) reveals that the shear modulus of BCC softens in a pressure range of 25-110 GPa, leading to the reduction of hardness in this region and the structure becomes more prone to plastic deformation. In this regard, the probability of a deviatoric stress-induced phase transformation into RH is high. The shear modulus of the three candidate structures satisfy a relation of RH2 > RH1 > BCC, and RH2 has the highest hardness. With the increasing shock pressure, the shear modulus of RH1 approaches that of BCC when beyond 175 GPa. In addition, we compare the theoretical data to the shear modulus estimated in Yu's experiment[12]. It is found that the calculated shear modulus of BCC is in good agreement with that of Yu. Since the Young's modulus depends on the bulk and shear modulus of the solid, the variation in Young's modulus is similar to that of shear modulus, as shown in Fig. S9(b). In both shear and Young's modulus, the softening with respect to the widely recognized SG model is very striking when pressure is below 100 GPa. On the other hand, when $P>100$ GPa, they harden much faster than the SG model, which is mainly contributed by the heating-induced hardening (HIH) effects of electrons, rather than the pure compression effect.

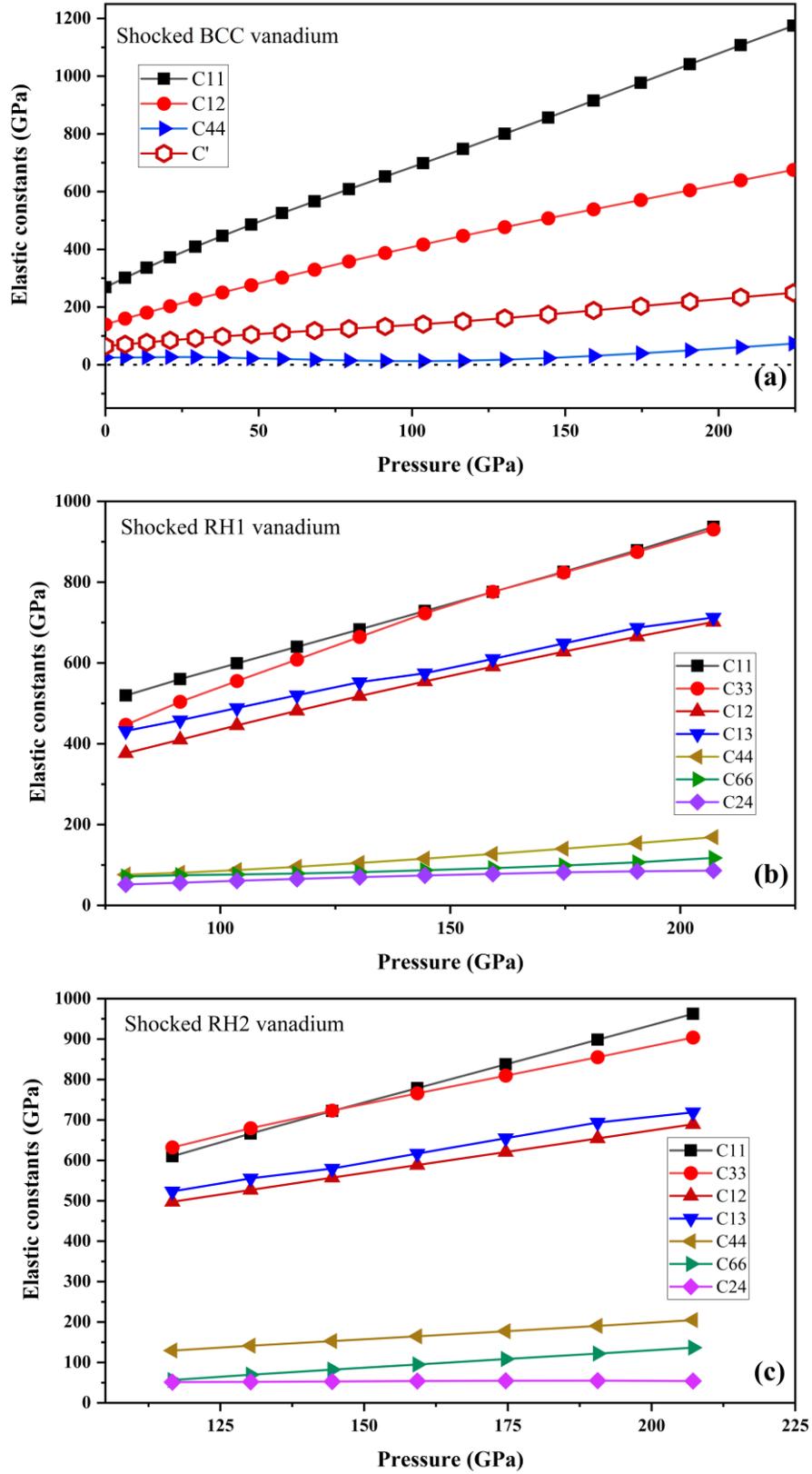

**Fig. S8** (Color online) Comparison of the calculated single-crystalline elastic constants of vanadium as a function of shock pressure in the candidate structures.

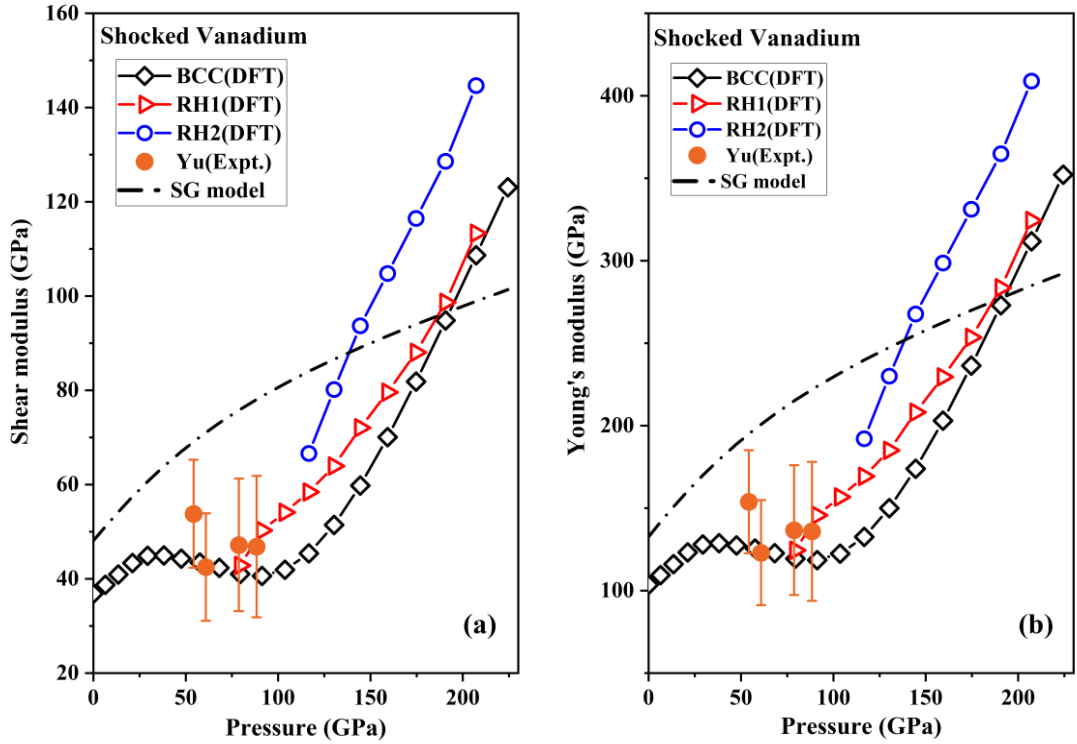

**Fig. S9** (Color online) Comparison of the polycrystalline elastic modulus of vanadium as a function of shock pressure in three candidate structures: (a) Shear modulus; (b) Young's modulus. The experimental estimates of Yu[12] are also displayed. The variation of shear and Young's modulus in the SG model[8, 13] along shock Hugoniot are given for comparison, respectively.

## 5. Steinberg–Guinan (SG) model for elasticity

Steinberg–Guinan (SG) model is a widely adopted phenomenological model to describe the variation of shear modulus with pressure and temperature. Using SG model, Rudd *et al.* derived the shear modulus variation of vanadium with pressure, which also includes the thermal lattice softening with temperature[8]. The SG model is given in a formula of[13]:

$$G(P,T) = G_0[1 + AP\eta^{-1/3} - B(T - 300)],$$

in which $G_0 B(T-300)$ is a correction term for the temperature softening due to thermal vibrations of the lattice. Figure S9 shows the shear and Young's modulus of SG model along the Hugoniot, which do not show a compression-induced softening (CIS) anomaly.

It is necessary to point out that the original correction term for temperature softening of the SG model cannot be applied to vanadium directly. From Walker's experiments[14], we knew that the lattice softening occurs only when $T > T_{max}$. At zero pressure, Walker et al. showed that $T_{max}$=1600 K (2500 K) for vanadium (niobium)[14], respectively. On the other hand, the melting temperature at ambient pressure is $T_m$=2183 K (2750 K) for vanadium (niobium). Walker's data indicates that the reduction rate of $C_{44}$ of vanadium should be $C_{44} \times 0.1 \times (T-T_{max})/1600$ when $T>T_{max}$. This relation can be extrapolated to the finite pressure, and the obtained reduction rate for the shear modulus is

$$\begin{cases} \dfrac{\Delta G}{G} = \dfrac{0.1}{1600} \times (T - 0.74 T_m) & , T > 0.74 T_m \\ \Delta G = 0 & , T < 0.74 T_m \end{cases}$$

On the other hand, according to Errandonea et al.'s recent experiment[15], the high pressure melting temperature of vanadium is about 3000 K (4000 K, 5000 K) at 50 GPa (100 GPa, 200 GPa), respectively. By Walker's estimate, the $T_{max}$ is thus about 2220 K (2960 K, 3700 K) according to $T_{max}$=0.74$T_m$, respectively. From the Hugoniot data as listed in Table S1, the Hugoniot temperature ($T_H$) of vanadium is 700 K (1500 K) at 50 GPa (100 GPa), respectively, which is much smaller than the $T_{max}$ at the same pressure. Therefore, the softening due to lattice vibrations as described by SG model can be safely ignored. At a higher pressure of 200 GPa, the $T_H$ is 3700 K, which is equal to $T_{max}$, and the lattice thermal softening can also be ignored.

This conclusion is in good agreement with early AIMD estimation[7], which showed that the lattice vibrational softening takes effect only when very close to the melting point. Because AIMD simulations always introduce large noise to the calculated energy, which is unwanted when to evaluate the smooth variation of elasticity with pressure and temperature, we therefore ignored this small but noisy lattice contribution in this work for the purpose of clarity, which will not affect our main results and conclusions.

## 6. The impact of core-core overlap

In our calculation, the 13 valence electrons ($3s^2 3p^6 3d^3 4s^2$) are considered. The PAW pseudopotential used in our calculation is called "V_sv" in VASP library. The core radius of the pseudopotential is 2.3 a.u. (1.217 Å), which will produce slight core-core overlap beyond about 75GPa. Therefore, in order to quantify the influence of this core-core overlap on the calculation results under high pressure, we use other two pseudopotentials with smaller core radius (and thus do not have core-core overlap problem) for comparison. The first is also PAW pseudopotential, which is called "V_sv_GW" in VASP library, and its core radius is 2.1 a.u. (1.111 Å). The second is USPP, which is generated on the fly in CASTEP software, and its core radius is 1.8 a.u. (0.9525 Å).

We use these three pseudopotentials to calculate the volume compression behavior and corresponding mechanical properties of BCC vanadium under static high pressures. The results are shown in Fig. S10. At about 200 GPa and 300 GPa, the nearest neighbor distances of vanadium atoms calculated by the three pseudopotentials are about 2.22 and 2.13 Å, respectively. It shows that the USPP does not have core-core overlap in a wider pressure range (0-300GPa). As shown in Fig. S10(a), the P-V relations calculated by the three pseudopotentials are almost identical. In sub-figures (b) to (i), we show the results of single-crystal elastic constants, polycrystalline modulus, and polycrystalline sound velocities, respectively. The results of the three pseudopotentials show good consistency, and the differences between them are also in a small range (less than ~5%).

We also compared the results under shock loading calculated by "V_sv" and "V_sv_GW" pseudopotentials, as shown in Fig.S11. There is no significant difference in polycrystalline modulus and polycrystalline sound velocity between the two pseudopotentials. For polycrystalline sound velocity, the maximum deviation between them under shock loading is less than 2%, which is good enough for our purpose in this work.

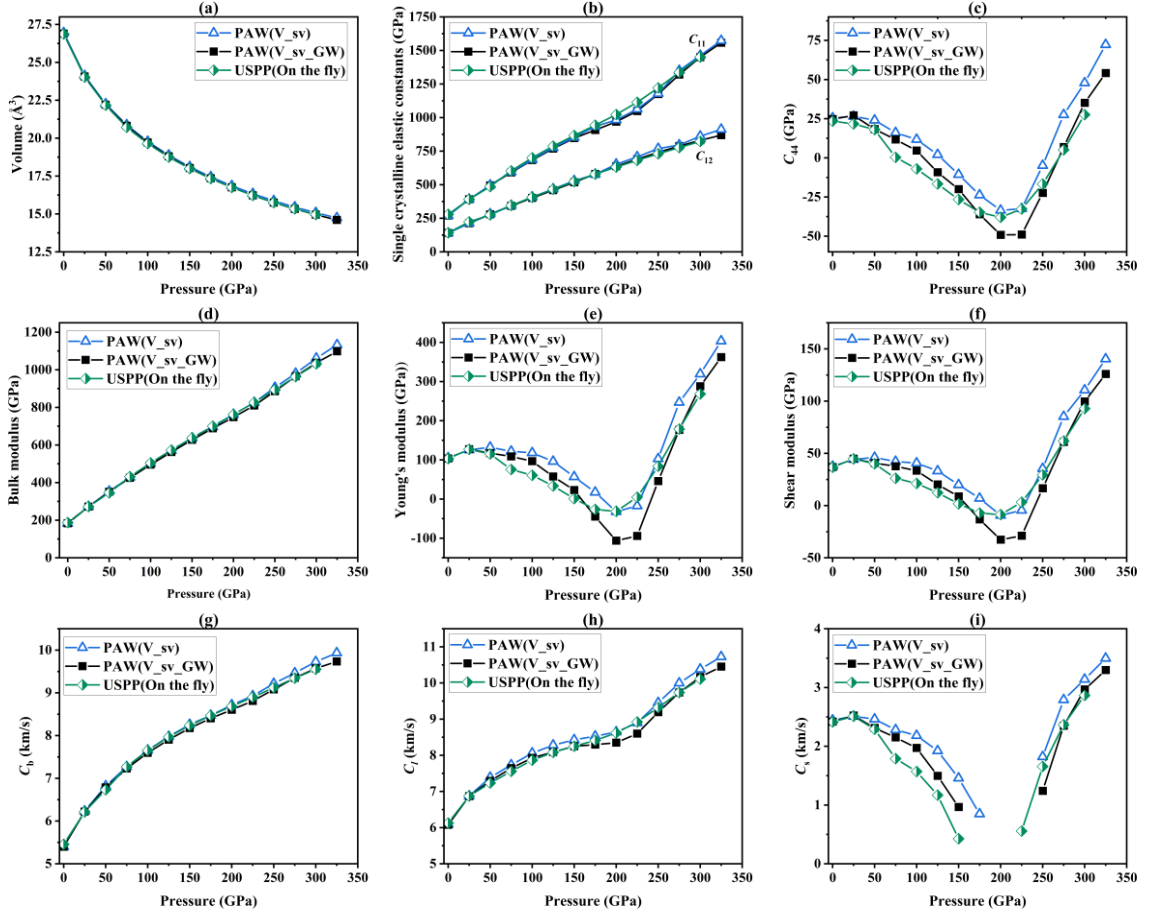

**Fig. S10** (Color online) Comparison of P-V relation (a), single crystal elastic constant (b-c), polycrystalline modulus (d-f) and polycrystalline sound velocity (g-i) calculated by three different pseudopotentials.

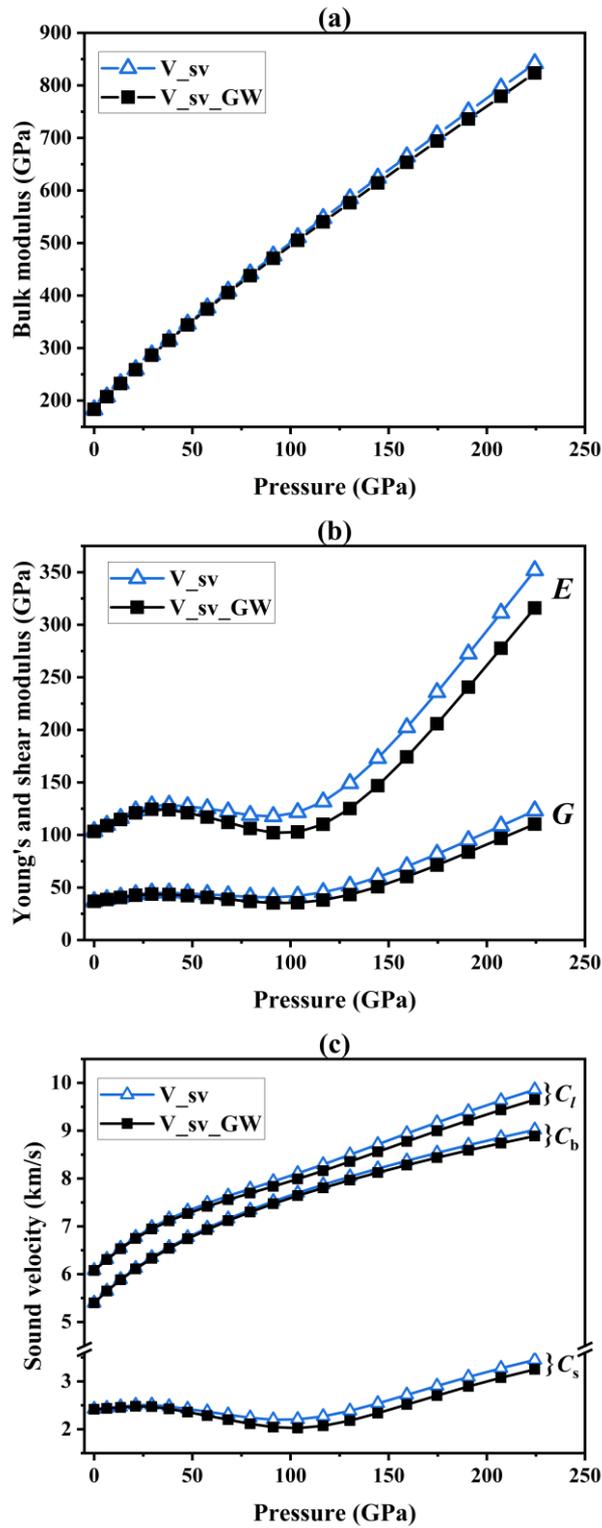

**Fig. S11** (Color online) Comparison of polycrystalline modulus (a-b) and polycrystalline sound velocity (c) calculated by V_sv and V_sv_GW pseudopotentials in shock compression.

In conclusion, for the PAW pseudopotential we used, the slight core-core overlap does not have a nonnegligible impact on the calculation results. In principle these core spheres should not overlap, but a bit of overlap/softness-of-the-potential trade-off is usually acceptable. Especially when spherical Bessel functions were used to construct the PAW potential, as implemented in VASP code, a large overlap between the atomic spheres is allowed[16, 17]. An extreme example of core-core overlap is hydrogen molecules at low pressures. The core radius of the VASP standard PAW pseudopotential for H is 0.58 Å, and the H-H distance in $H_2$ molecule is 0.74 Å. The pseudopotential thus has nearly 60% overlap, but this *huge* overlap hardly affects its applications in hydrogen, as documented in many hydrogen-related literatures.